\documentclass{emulateapj}

\usepackage{graphicx,natbib,amssymb}
\usepackage{lscape}

\def\astrobj#1{#1}

\shorttitle{Analysis of 20 eclipsing binary light-curves observed by the INTEGRAL/OMC}
\shortauthors{P.Zasche}

\begin{document}

\title{The data mining: An analysis of 20 eclipsing binary light-curves observed by the INTEGRAL/OMC}

\author{P. Zasche\footnote{eMail to: zasche@sirrah.troja.mff.cuni.cz}}
\address{Astronomical Institute, Faculty of Mathematics and Physics,
 Charles University Prague, CZ-180 00 Praha 8, V Hole\v{s}ovi\v{c}k\'ach 2, Czech Republic}

\begin{abstract}
 Twenty eclipsing binaries were selected for an analysis from a huge database of observations made
 by the INTEGRAL/OMC camera. The photometric data were processed and analyzed, resulting in a first
 light-curve study of these neglected eclipsing binaries. Most of the selected systems are the
 detached ones. The system ET~Vel was discovered to be an eccentric one. Due to missing
 spectroscopic study of these stars, further detailed analyses are still needed.
\end{abstract}


\section{Introduction}

The INTEGRAL (INTErnational Gamma-Ray Astrophysics Laboratory) satellite was launched on 17 October
2002. Since then, there were many observations of the gamma-ray and X-ray sources obtained. Thanks
to the OMC (Optical Monitoring Camera), there was collected also a large database of the
photometric observations in the visual passband. The variable stars observations were obtained as a
by-product.

A huge database of all observations obtained by OMC is freely available on internet, but the
analyses are still very rare. The present paper is following a similar study of three selected
Algol-type binaries \astrobj{PV Cyg}, \astrobj{V822 Cyg}, and \astrobj{V1011 Cyg}, see
\cite{Zasche2008NewA}. The selection criteria used here were the following: Maximum number of data
points and non-existence of any light-curve analysis of the particular system. There were 20
systems selected for the present paper.

\section{Analysis of the individual systems}

All observations of these systems were carried out by the same instrument (50mm OMC telescope) and
the same filter (Johnson's V filter). Time span of the observations ranges from November 2002 to
July 2006. A transformation of the time scale has been done following the equation $Julian Date -
ISDC Julian Date = 2,451,544.5$. Only a few outliers from each data set were excluded. The {\sc
Phoebe} programme (see e.g. \citealt{Prsa2005}), based on the Wilson-Devinney algorithm
\citep{Wilson1971}, was used.

Due to missing information about the stars, and having only the light curves in one filter, some of
the parameters have to be fixed. At first, the "Detached binary" mode was used for computing
(except for V435~Cas). The limb-darkening coefficients were interpolated from van~Hamme's tables
(see \citealt{vanHamme1993}), the linear cosine law was used. The values of the gravity brightening
and bolometric albedo coefficients were set at their suggested values for convective atmospheres
(see \citealt{Lucy1968}), i.e. $G_1 = G_2 = 0.32$, $A_1 = A_2 = 0.5$. In all cases (except for
ET~Vel) the orbital eccentricity was set to 0 (circular orbit). Therefore, the quantities which
could be directly calculated from the light curve are the following: The luminosity ratio
$L_1/L_2$, the temperature ratio $T_1/T_2$, the inclination $i$, ephemerides of the system, the
Kopal's modified potentials $\Omega_1$ and $\Omega_2$, the synchronicity parameters $F_1$ and
$F_2$, the third light $l_3$, and the mass ratio $q$. Because we deal with the detached systems,
the last quantity was derived via a "q-search method", which means trying to find the best fit with
different values of $q$ ranging from 0 to 1 with a step 0.1 (all systems except for V466 Cyg range
in these limits). Using the parameters introduced above, one could also derive the value of the
radii ratio $R_1/R_2$.

The distinguishing between the minima has been done only according to the observational point of
view, which means that the deeper one is the primary one. This results in a fact that the primary
component could be neither the larger one, nor the more massive one. In a few cases the secondary
components result to be the more luminous ones, and in one case (V466~Cyg) also the more massive
one.

All of the basic information about the analyzed systems are introduced in Table \ref{Table1}, where
are the magnitudes from the GCVS \citep{1971GCVS} compared with the actual OMC magnitudes in
Johnson's $V$ filter, the depths of both primary and also secondary minima in $V$ filter, the
orbital periods and also the $B-V$ indices (mostly taken from the NOMAD catalogue,
\cite{NOMAD2004}, which contains the most complete $B$ and $V$ observations of these stars). The
spectral types were taken from various sources (see SIMBAD database).

\begin{table*}[t!]
\small \caption{Basic information about the analyzed systems.}
 \label{Table1} \centering \scalebox{0.90}{
\begin{tabular}{c c c c c c c c }
\hline
    Star   & Mag$_{GCVS}$ & Mag$_{OMC}$ & Mag$_{MinI}$ & Mag$_{MinII}$ & Period (d) & $B-V$ & Sp.\\
    \hline
    ZZ Cas &    10.80     &    10.87    &    11.49     &     11.19     &   1.2435   & 0.43  & B3 \\
    BG Cas &    12.90     &    12.79    &    13.68     &     12.90     &   3.9537   & 0.23  &    \\
    EN Cas &    12.30     &    11.39    &    12.22     &     11.66     &   4.4378   & 0.47  &    \\
    GT Cas &    11.90     &    11.55    &    12.37     &     11.59     &   2.9899   & 0.29  & A0 \\
    IL Cas &    11.50     &    11.09    &    11.71     &     11.18     &   3.4517   & 0.19  & B5 \\
  V435 Cas &    15.00     &    13.85    &    15.06     &     14.19     &   4.1561   & 0.90  &    \\
    CM Cep &    12.10     &    11.88    &    14.07     &     12.02     &   1.8589   & 0.36  & B8 \\
    DP Cep &    12.60     &    12.86    &    14.13     &     13.12     &   1.2700   & 0.22  &    \\
    SZ Cru &    10.90     &    11.49    &    12.45     &     11.58     &   1.9743   & 0.25  &    \\
    AY Cru &    11.10     &    11.28    &    12.24     &     11.38     &   1.5984   & 0.37  &    \\
    GG Cyg &    12.10     &    11.93    &    12.81     &     12.08     &   2.0084   & 0.38  & A4 \\
  V466 Cyg &    10.80     &    10.56    &    11.30     &     11.16     &   1.3916   & 0.40  & A8 \\
  V809 Cyg &    13.30     &    13.10    &    13.85     &     13.53     &   1.9645   & 0.11  & F9 \\
     Z Nor &     9.30     &     9.15    &    10.09     &      9.47     &   2.5569   & 0.15  & B3IV \\
 V1001 Oph &    14.40     &    13.97    &    15.40     &     14.20     &   1.7894   & 0.96  &    \\
  V986 Sgr &    13.00     &    12.61    &    13.81     &     12.72     &  10.4297   & 0.58  &    \\
    BY Vel &    10.60     &    11.01    &    12.11     &     11.09     &   3.4553   & 0.50  &    \\
    CZ Vel &    10.80     &    10.83    &    11.50     &     11.05     &   5.1927   & 0.50  &    \\
    DF Vel &    13.30     &    13.33    &    15.11     &     13.64     &   0.7645   & 0.48  &    \\
    ET Vel &    11.90     &    11.14    &    11.76     &     11.75     &   3.0809   & 0.50  & A0 \\ \hline
 \hline
\end{tabular}}
\end{table*}

\begin{figure}[b]
 \hskip -1cm
 \includegraphics[width=19.5cm]{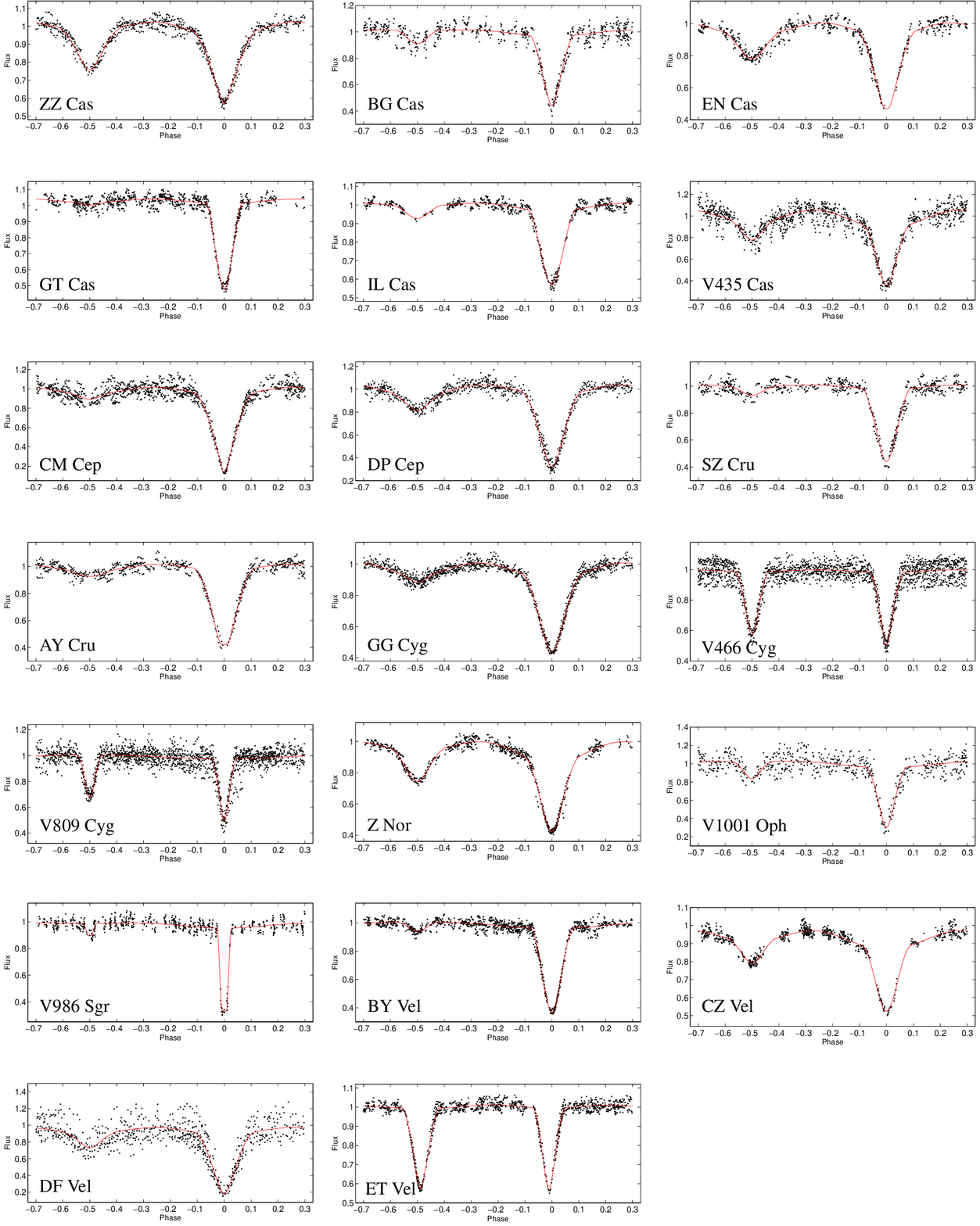}
 \caption{The light curves of the analyzed systems.}
 \label{Figs}
\end{figure}

The results are introduced in Fig.\ref{Figs} and Table \ref{Table2}, where are given all relevant
parameters of the analyzed systems. Inclinations smaller than $90^\circ$ mean that the binary
rotates counter-clockwise as projected onto a plane of sky. Only one system (DF~Vel) has its
orbital period shorter than 1~day, V435~Cas is the only semi-detached system, and ET~Vel was found
to be the eccentric eclipsing binary. Its parameters are the following: the eccentricity $e =
0.0737$ and the argument of periastron $\omega = 1.067$~rad. In this system the temperature of the
primary component (and also its luminosity) is lower than temperature of the secondary one, we can
therefore doubt about the role of the two components. Moreover, both primary and secondary minima
have approximately equal depths, so the primary and secondary components are probably interchanged.

Another interesting fact of this sample are the relative radii of these stars. In about one half of
the systems the secondary components are greater than the primaries. Concerning the luminosities,
there are also 5 cases where the third light (from the unseen component) has a statistically
significant value above 5\%. One could speculate about a prospective future discovery of such
components in these systems. Due to missing detailed analysis (spectroscopic, interferometric,
etc.), the only possible way how to discover these bodies nowadays is the period analysis of their
times of minima variations. In the system ZZ~Cas such an analysis exists and the third body was
discovered, see \cite{ZZCas1991BAICz}.

\section{Discussion and conclusions}

The light-curve analyses of twenty selected systems have been carried out. Using the light curves
observed by the INTEGRAL satellite, one can estimate the basic physical parameters of these
systems. Despite this fact, the parameters are still only the preliminary ones, affected by
relatively large errors and some of the relevant parameters were fixed at their suggested values.
The detailed analysis is still needed, especially in different filters. Together with a prospective
radial-velocity study, the final picture of these systems could be done. Particularly, the system
ET~Vel seems to be the most interesting one due to its eccentric orbit.

\section{Acknowledgments}
Based on data from the OMC Archive at LAEFF, pre-processed by ISDC. This investigation was supported by the
Grant Agency of the Czech Republic, grants No. 205/06/0217 and No. 205/06/0304. We also acknowledge the support
from the Research Program MSMT 0021620860 of the Ministry of Education. This research has made use of the SIMBAD
database, operated at CDS, Strasbourg, France, and of NASA's Astrophysics Data System Bibliographic Services.

%\begin{landscape}
\begin{table*}
\small \caption{The light-curve parameters of the individual systems.}
 \label{Table2} \centering \scalebox{0.87}{
\begin{tabular}{c c c c c c c c c c c c c c c c }
\hline
 Parameter &  HJD$_0$ &       P     &   $i$  &    $q$     & $\Omega_1$ & $\Omega_2$ & $T_1/T_2$ & $L_1$ & $L_2$ & $L_3$ & $R_1/R_2$ & $F_1$ & $F_2$ & $x_1$ & $x_2$ \\
    Star   & 2452000+ &    [days]   &  [deg] & $=M_2/M_1$ &            &            &          & [\%]  & [\%]  & [\%]  &           &       &       &       &       \\
    \hline
    ZZ Cas & 637.564  &  1.24349377 & 86.552 &    0.7     &    3.664   &    3.583   &   1.49   &  55.0 & 21.4  & 23.6  &   1.17    & 1.439 & 1.560 & 0.287 & 0.367 \\
    BG Cas & 655.132  &  3.95367825 & 96.126 &    0.5     &    4.153   &    3.926   &   2.02   &  86.4 &  6.3  &  7.3  &   1.32    & 1.400 & 3.496 & 0.412 & 0.718 \\
    EN Cas & 657.014  &  4.43777896 & 86.727 &    0.6     &    3.492   &    3.613   &   1.07   &  80.5 & 15.5  &  4.0  &   1.33    & 1.159 & 2.082 & 0.612 & 0.668 \\
    GT Cas & 641.291  &  2.98985074 & 78.687 &    0.2     &    6.143   &    2.141   &   2.54   &  81.0 &  7.0  & 12.0  &   0.66    & 5.155 & 0.000 & 0.248 & 0.422 \\
    IL Cas & 639.814  &  3.45171726 & 75.645 &    0.6     &    4.224   &    3.195   &   1.43   &  80.2 & 11.2  &  8.6  &   0.92    & 0.414 & 0.000 & 0.504 & 0.488 \\
  V435 Cas & 656.806  &  4.15607055 & 81.598 &    0.6     &    6.060   &    3.187   &   1.46   &  64.7 & 31.7  &  3.6  &   0.68    & 5.458 & 1.242 & 0.852 & 0.500 \\
    CM Cep & 654.610  &  1.85892148 & 90.305 &    0.4     &    3.873   &    2.650   &   1.65   &  93.0 &  5.4  &  1.6  &   1.04    & 2.599 & 1.008 & 0.507 & 0.852 \\
    DP Cep & 655.170  &  1.26999723 & 83.405 &    0.7     &    4.122   &    3.283   &   1.22   &  75.7 & 19.3  &  5.0  &   0.90    & 1.715 & 1.028 & 0.732 & 0.828 \\
    SZ Cru & 702.564  &  1.97430610 & 82.118 &    0.6     &    4.150   &    3.534   &   2.32   &  89.6 &  6.7  &  3.7  &   1.09    & 0.889 & 0.925 & 0.087 & 0.276 \\
    AY Cru & 702.551  &  1.59841897 & 90.944 &    0.4     &    3.154   &    2.780   &   1.27   &  95.6 &  4.1  &  0.3  &   1.34    & 0.810 & 0.818 & 0.458 & 0.527 \\
    GG Cyg & 595.973  &  2.00837615 & 76.705 &    0.2     &    3.204   &    2.084   &   2.83   &  76.0 &  7.0  & 17.0  &   1.19    & 1.505 & 0.484 & 0.285 & 0.530 \\
  V466 Cyg & 595.223  &  1.39156321 & 89.365 &    1.1     &    6.106   &    7.060   &   1.09   &  58.8 & 40.6  &  0.6  &   1.13    & 1.923 & 0.000 & 0.523 & 0.575 \\
  V809 Cyg & 596.231  &  1.96445344 & 92.850 &    0.6     &    8.095   &    5.959   &   1.73   &  59.0 & 38.6  &  2.4  &   1.01    & 5.888 & 6.292 & 0.425 & 0.629 \\
     Z Nor & 669.153  &  2.55694871 & 81.435 &    0.7     &    4.130   &    3.818   &   1.47   &  73.1 & 23.2  &  3.7  &   1.06    & 1.861 & 2.188 & 0.818 & 0.500 \\
 V1001 Oph & 787.850  &  1.78942143 & 86.958 &    0.6     &    5.049   &    4.238   &   1.85   &  78.3 & 19.4  &  2.3  &   0.99    & 0.000 & 3.541 & 0.611 & 0.500 \\
  V986 Sgr & 730.524  & 10.42963578 & 87.746 &    0.4     &   22.835   &    5.341   &   1.94   &  33.7 & 64.8  &  1.5  &   0.39    & 0.210 & 8.807 & 1.000 & 0.500 \\
    BY Vel & 650.590  &  3.45533443 & 80.704 &    0.4     &    5.308   &    2.817   &   2.61   &  88.0 &  8.8  &  3.2  &   0.80    & 3.514 & 0.968 & 0.274 & 0.504 \\
    CZ Vel & 651.500  &  5.19274753 & 79.017 &    0.4     &    4.604   &    2.822   &   1.35   &  76.0 & 23.9  &  0.1  &   0.94    & 3.376 & 1.426 & 0.500 & 0.500 \\
    DF Vel & 650.762  &  0.76447904 & 87.432 &    0.9     &    4.154   &    3.691   &   1.28   &  78.4 & 21.4  &  0.2  &   0.89    & 0.579 & 0.691 & 0.751 & 0.657 \\
    ET Vel & 749.273  &  3.08087500 & 86.663 &    0.9     &    6.131   &    5.615   &   0.96   &  47.4 & 51.6  &  1.0  &   0.97    & 1.032 & 1.000 & 0.680 & 0.646 \\ \hline
 \hline
\end{tabular}}
\end{table*}
%\end{landscape}

\begin{thebibliography}{}

 \bibitem[Kreiner \& Tremko(1991)]{ZZCas1991BAICz} Kreiner, J.~M., \& Tremko, J.\ 1991, Bulletin of
 the Astronomical Institutes of Czechoslovakia, 42, 345

 \bibitem[Kukarkin et al.(1971)]{1971GCVS} Kukarkin, B.~V., Kholopov, P.~N., Pskovsky, Y.~P., Efremov, Y.~N., Kukarkina, N.~P.,
 Kurochkin, N.~E., \& Medvedeva, G.~I.\ 1971, General Catalogue of Variable Stars, 3rd ed.~(1971)

 \bibitem[Lucy(1968)]{Lucy1968} Lucy, L.~B.\ 1968, \apj, 151, 1123

 \bibitem[Pr{\v s}a \& Zwitter(2005)]{Prsa2005} Pr{\v s}a, A., \& Zwitter, T.\ 2005, \apj, 628, 426

 \bibitem[van Hamme(1993)]{vanHamme1993} van Hamme, W.\ 1993, \aj, 106, 2096

 \bibitem[Wilson \& Devinney(1971)]{Wilson1971} Wilson, R.~E., \& Devinney, E.~J.\ 1971, \apj, 166, 605

 \bibitem[Zacharias et al.(2004)]{NOMAD2004} Zacharias, N., Monet, D.~G., Levine, S.~E., Urban, S.~E., Gaume, R., \&
 Wycoff, G.~L.\ 2004, Bulletin of the American Astronomical Society, 36, 1418

 \bibitem[Zasche(2008)]{Zasche2008NewA} Zasche, P.\ 2008, New Astronomy, 13, 481

\end{thebibliography}
\end{document}